\documentclass[]{elsarticle}

\usepackage{hyperref}
\usepackage[utf8]{inputenc}
\usepackage[english]{babel}
\usepackage{multirow} %Mehr-zeilige/-spaltige Tabelle
\usepackage{float} % Fixe Position
\usepackage{textcomp}
\usepackage{booktabs}
\usepackage{amsmath}
\usepackage{units}
\usepackage{color,soul} 
\usepackage{ifpdf} 
\ifpdf % if using pdfLaTeX in PDF mode
\usepackage{graphicx} 
\usepackage{pgfplots} 
\newlength\figureheight
\newlength\figurewidth
\usetikzlibrary{shapes,arrows,backgrounds,fit,positioning,chains,calc,automata,matrix,trees,decorations,patterns}
\usetikzlibrary{decorations.pathreplacing, decorations, positioning, calc, patterns, arrows, math, shadings, arrows.meta}
\else %if using LaTeX or pdfLaTeX in DVI mode 
\usepackage{graphicx} 
\DeclareGraphicsExtensions{.eps,.bmp} 
\usepackage{pgf} 
\usepackage{tikz} 
\usepackage{caption}
\usepackage{subcaption}
\fi

\usepackage{listings}
\lstdefinestyle{myxml}{
	belowcaptionskip=5pt,
	abovecaptionskip=0pt,
	belowskip = 4pt,
	aboveskip = 4pt,
	breaklines=true,
	frame=bt,
	keywords=[1]{nodeTypes, nodeType, layerTypes, layerType, nodes, node, connections, con, ports, port, map, bind, data, dataTypes, dataType, task, requires, requirement, generates, possibility, destinations, destination, synthetic, phase, distribution },
	keywords=[2]{routerModel, routing, selection, arbitration, clockSpeed, technology, xPos, yPos, zPos, idType, length, width, depth, interface, bufferDepth, vcCount, name, start, duration, repeat, type, souce, count, probability, delay, interval, count, source, hotspot},
	keywords=[3]{id, value, min, max},
	showstringspaces=false,
	basicstyle=\footnotesize\ttfamily,
	keywordstyle=[1]\color{col1},
	keywordstyle=[2]\color{col2},
	keywordstyle=[3]\color{col3},
	commentstyle=\itshape\color{green},
	stringstyle=\color{purple},
	identifierstyle=\color{gray},
	tabsize=2,
	numbersep=8pt,
	stepnumber=1,
	numberstyle=\tiny\color{gray},
}

\pgfplotsset{compat=newest}
\usepackage{pgfplotstable}
\usepgfplotslibrary{groupplots}
\pgfplotsset{scaled y ticks=false} % turn of scientif notations in y ticks

\usetikzlibrary{fadings}

\tikzset{router/.style={circle,draw,fill=gray!60,inner sep=0pt,minimum size=5pt}}
\tikzset{desc/.style={font = \scriptsize}}
\tikzset{cross/.style={cross out, draw, 
		minimum size=2*(#1-\pgflinewidth), 
		inner sep=0pt, outer sep=0pt}}

\definecolor{col1}{RGB}{27,158,119}
\definecolor{col2}{RGB}{217,95,2}
\definecolor{col3}{RGB}{117,112,179}

\makeatletter
\renewcommand{\fnum@figure}{Fig.~\thefigure}
\makeatother

\journal{Integration}

\begin{document}

\begin{frontmatter}

\title{Simulation Environment for Link Energy Estimation in Networks-on-Chip with Virtual Channels}

	\author[label1]{Jan Moritz Joseph\fnref{myfootnote}}%\corref{cor1}}
	\ead{jan.joseph@ovgu.de}
	\ead[url]{www.iikt.ovgu.de}
	\fntext[myfootnote]{Authors contributed equally.}
	
	\author[label2]{Lennart Bamberg\fnref{myfootnote}}%\corref{cor1}}
	\ead{bamberg@item.uni-bremen.de}
	\ead[url]{www.item.uni-bremen.de}
	
	\author[label1]{Imad Hajjar}%\corref{cor1}}
	\ead{imad.hajjar@st.ovgu.de}
	\ead[url]{www.iikt.ovgu.de}
	
	\author[label2]{Robert Schmidt}%\corref{cor1}}
	\ead{rschmidt@item.uni-bremen.de}
	\ead[url]{www.item.uni-bremen.de}
	
	\author[label1]{Thilo Pionteck}
	\ead{thilo.pionteck@ovgu.de}
	\ead[url]{www.iikt.ovgu.de}
	
	\author[label2]{Alberto Garc\'ia-Ortiz}
	\ead{agarcia@item.uni-bremen.de}
	\ead[url]{www.item.uni-bremen.de}
	
	\address[label1]{Otto-von-Guericke-Universit\"at Magdeburg\\
		Institut f\"ur Informations- und Kommunikationstechnik (IIKT)\\
		39106 Magdeburg, Germany}
	\address[label2]{University of Bremen\\
		Institute of Electrodynamics and Microelectronics (ITEM.ids)\\
		28359 Bremen, Germany}

\begin{abstract}
Network-on-chip (NoC) is the most promising design paradigm for the interconnect architecture of a multiprocessor system-on-chip (MPSoC).
On the downside, a NoC has a significant impact on the overall energy consumption of the system.
NoC simulators are highly relevant for design space exploration even at an early stage. Since links in NoC consume up to \unit[50]{\%} of the energy, a realistic energy consumption of links in NoC simulators is important. 
This work presents a simulation environment which implements a technique to precisely estimate the data dependent link energy consumption in NoCs with virtual channels for the first time. Our model works at a high level of abstraction, making it feasible to estimate the energy requirements at an early design stage.
Additionally, it enables the fast evaluation and early exploration of low-power coding techniques.
The presented model is applicable for 2D and 3D NoCs. A case study for an image processing application shows that the current link model leads to an underestimate of the link energy consumption by up to a factor of four. In contrast, the technique presented in this paper estimates the energy quantities precisely with an error below \unit[1]{\%} compared to results obtained by precise, but computational extensive, bit-level simulation.

\end{abstract}


\end{frontmatter}

\section{Introduction}
Packet-switching-based networks-on-chip (NoC) build the most promising interconnect architecture for multi-core systems-on-chip (SoC)~\cite{de2006networks}. However, the link's energy consumption of a NoC has increased substantially with the ongoing scaling of technology~\cite{jafarzadeh2014data}. A model to estimate the energy requirements at high abstraction levels is important to determine the required energy budget for the links and to evaluate the effect of high-level energy optimization approaches. Lower-level models are not optimal due to their long runtime. Furthermore, high-level optimization approaches have a significantly higher impact on the energy requirements than low-level ones~\cite{Eisley:2004}. 

However, the current link model~\cite{jafarzadeh2014data,5752410} is only valid for a sequential transmission of single packets, which is not the case when virtual channels are used. Virtual channels are a commonly implemented technique to increase the NoC performance~\cite{de2006networks}, but they also lead to drastic increases in switching activities, as shown in this work. Since switching probabilities are directly proportional to energy consumption, neglecting the effect of virtual channels leads to a heavy underestimation of energy requirements. 

Also, virtual channels have a major impact on coding, which is one of the most promising low-power approaches.
Since the overhead of a coding approach must not cancel out savings in links, encoding is typically based on an end-to-end manner~\cite{de2006networks,jafarzadeh2014data}, in which data is encoded and decoded at the network interfaces of the source and destination respectively, and not on link level. However, on an end-to-end basis the effect of virtual channels can not yet be considered and the energy reduction of most techniques vanishes when virtual channels are used. Hence, the majority of techniques are explicitly designed for NoCs without virtual channels~\cite{jafarzadeh2014data,5752410}, even though they are commonly implemented in NoCs.
The only technique which is applicable for virtual-channel-based NoCs is probability coding~\cite{garcia2010practical}. However, this method was only designed for the uncommon scenario that links permanently make use of non-prioritized virtual channels. Thus, no optimal coding approach can be identified yet, due to the lack of a pattern-dependent model for the energy consumption of links with virtual channels. Furthermore, such a model is required to design efficient coding approaches, as outlined in this work. 

This work is an invited extension of Ref.~\cite{Bamberg.2018}. In this initial work, \emph{Bamberg et al.} proposed the first accurate, coding-aware, high-level model for the energy consumption of 2D and 3D links with virtual channels. In detail, the existing link energy model for 2D links~\cite{jafarzadeh2014data,5752410} and 3D links~\cite{7961256} is extended in a way that it enables estimating the switching characteristics not only for a sequential transmission of data packets (no virtual channels), but also for multiplexing individual flits of multiple packets (virtual channels). While the existing energy model likely results in implementing inefficient coding techniques and underestimates the link's energy requirements by up to a factor of four, the model introduced precisely predicts the energy requirements of 2D and 3D links, as well as coding efficiencies. We achieve a power estimation error below \unit[1]{\%} compared to results obtained by precise, but computational extensive, bit-level simulation. Thereby, the proposed model reveals new options to reduce the link energy consumption using coding. 

The present work provides the following major extensions over the initial conference publication~\cite{Bamberg.2018}: (a) We make the aforementioned models available and usable for NoC simulations. Precise estimation of the energy budget is essential because links participate a large portion of the overall NoC energy consumption. Evidence therefore are a \unit[17]{\%} share in the Teraflop router \cite{Kahng:2009} and share of up to \unit[53.9]{\%} in the NOSTRUM $8\!\times\!8$-NoC \cite[Table 1, p. 4]{Penolazzi.2006}. Our NoC simulator comprises a NoC and an application model beside the proposed energy models. The simulator is implemented in C++ using the SystemC class library \cite{SystemCStandard} including both models on the transaction and the cycle-accurate level, while the energy models are implemented in Python. Our approach enables separate consideration of TSV arrays and of metal wires in NoCs, as well as accounting for application-dependent switching activity for individual links. This is orthogonal to energy estimations provided by competing simulators that do not implement such energy and application models. (b) We introduce the innovative feature of transmission matrices that are recorded during simulation. Thereby, parts of the architectural design can be modified post-simulation. For instance, the evaluation of various coding methods can be conducted only with a single simulation run. (c) Our models are easily configurable using XML files, including many properties of the architecture of individual routers. This approach facilitates rapid prototyping during design space exploration. (d) We enable an easy use of the simulator by providing a single point-of-entry and simply configurable simulation scripts. Summarizing the novelty of this paper: We improve state-of-the-art energy estimation for NoCs such as \cite{kahng2015orion3} by accounting for capacitance, separation between metal wires and TSV arrays, and coding. We demonstrate applicability of our approach by comparison of virtual channel-based NoCs and NoCs without virtual channels in a realistic case study and show power and performance as well as the effects of coding.

The remainder of this work is structured as follows: Sec.~\ref{Sec1} presents general formulas to determine the link energy consumption employing the switching/bit properties. Sec.~\ref{sec:eff} reviews the concept of virtual channels and shows that they have a huge impact on the switching properties. Afterwards, our model to estimate the switching characteristic is presented in Sec.~\ref{sec:model}. The model is integrated into a NoC simulator, which is presented in Sec.~\ref{sec:simulator}. Further, we validate the model by means of simulation results in Sec.~\ref{sec:sim_res}. In Sec.~\ref{sec:cs}, a case study for a heterogeneous 3D vision SoC is presented which shows that for a precise energy estimation, as well as a design of efficient coding architectures, our model has to be used. Finally, a conclusion is drawn.

\section{Related Work}

Due to increasing need for power efficient systems, low-power link encoding has drawn a lot of attention in recent years (e.g.~\cite{jafarzadeh2014data,5752410, GarciaOrtiz.2017, GarciaOrtiz.2011}). To estimate the pattern-dependent link energy consumption, and thereby the coding efficiency, as a function of the network architecture and the application specific data flow, NoC simulators are commonly used \cite{Catania.2016}. However, current link energy models do not consider the effect of virtual channels, even though they are usually implemented to increase the NoC performance. This leads to a lack of efficient coding approaches applicable for virtual channel-based NoCs. 

%As already introduced, there are two energy models for links, namely \cite{jafarzadeh2014data,5752410}. \emph{Jafarzadeh et al.} \cite{jafarzadeh2014data} contribute three different encoding schemes for NoCs, which decode and encode the data in the network interface before injection into the network, i.e.\ end-to-end coding. The authors achieve up to 51\% saved power dissipation and up to 14\% reduced energy consumption. There is no significant performance penalty and the hardware implementation shows only 15\% area overhead in the network interface. \emph{Palesi et al.} \cite{5752410} also proposes an end-to-end encoder in the network interface, which is integrated into the network interface. While it achieves power dissipation savings of 37\% and a declined energy consumption of 18\%, it has no significant performance or area costs. Both publications do only consider sequential transmission of flits from a single packet on links to calculate the energy and power results. As already stated, this does not cover the relevant functionality of NoCs, which use virtual channels: These switch between flits of different packets on a regular basis. Thus, the assumption of sequential transmission does not hold for NoCs, in general. Therefore, the contribution of this work is orthogonal to the related work.

Generally, there exist a wide range of software for the simulation of NoCs at high abstraction levels, which can be divided into two classes: First class, simulators which target a specific architecture or use case, and the second class in which they are universally applicable. The first category of simulators is usually proposed as a supplementary to novel NoC hardware designs. One example is \cite{Chao.2010}, which is a simulator written for a specific 3D router architecture; in \cite{Joseph.2016}, a router architecture which accelerates data streams in NoCs is proposed with a simulator implementing this specific router model. In \cite{Park.2008}, a NoC router is extended over multiple layers in a 3D chip and an application specific simulator is proposed. However, all the aforementioned examples have limited applicability. 

In the second category, simulators target as many architectures, designs and use cases as possible. There are two well-maintained and well-known software: \emph{Noxim} \cite{Catania.2016} simulates NoCs on cycle-accurate abstraction level and is implemented in C++ using SystemC. To cover multiple use cases, many parameters of the architecture can be freely defined (e.g.\ buffer depth, packet size, routing algorithm, network topology). The simulator returns performance and energy figures for the NoC. The power is calculated using  a simple, cycle-accurate, event-based model. In Noxim, a fixed energy consumption is assigned to events. These are counted and their energy consumption is accumulated to a static value. Among others, one event is the transmission of a flit over a link. However, we will show in this work that the energy consumption of a link transmitting a flit strongly depends on the switching allocation, associated with the virtual channels and not only on the accumulated number of transmitted flits. The simulator \emph{BookSim 2.0} \cite{Jiang.2013} is similar; it is also cycle-accurate, implemented in C++ and offers similar parameters to set. It does not offer detailed power analysis and only provides network statistics. 

There are many works for power estimation in NoC simulations. ORION 3.0 \cite{kahng2015orion3} provides the most detailed models for energy consumption of the router's components. Therefore, it is currently state-of-the-art to estimate router power. It also includes a basic power model for links, which is introduced in Eqs. 10 and 11 in Ref.~\cite{kahng2015orion3}. It does not account for the effects of pattern-dependent coupling switching, which leads to a modeling error of up to 79.77\% \cite{Bamberg.2017}. The authors themselves are aware of this limitation: They argue that their approach does not model power on flit level and that they "do not consider bit encodings in a flit, which can lead to significant errors in dynamic power estimation" \cite[Sec.~IV-C, p. 9]{orion3Tecnical}. Furthermore, our model allows analyzing any geometrical TSV or metal wire dimension an array sizes and thus can be adapted to all technologies, including heterogeneous 3D integration. This is not accounted for by ORION 3.0. Within Ref.\ \cite{orion3Tecnical}, the authors propose to use a rather old NoC simulator GARNET \cite{4919636} from 2009 for full-system NoC simulation to overcome limitations of ORION 3.0. This would allow modeling flit-level power. We also extend this approach because evaluation of different codings is possible post-simulation without the necessity of a slow full-system simulation. To summarize, the manuscript submitted is the first work to account for physical effects in links on flit-level including pattern-dependent coupling and coding and including. 

To summarize, the proposed NoC simulator extends the existing solutions. It peruses a general approach and, as a unique feature, implements the proposed energy model to estimate the dynamic link energy. Furthermore, it is based on well-defined models, as proposed in \cite{Joseph.2018} and offers a structured design process for 3D NoC targeting heterogeneous 3D integrated circuits (ICs) as contributed in \cite{Joseph.2016c}.
\section{Data dependent link energy consumption}\label{Sec1}
The mean, pattern dependent, energy consumption of an $N$-bit link can be precisely estimated using~\cite{7961256,5876688}:
%\begin{equation} \label{eq:P_coup}
%{P}=\frac{V_{dd}^2f}{2}\left[\sum_{i=1}^{N}\mathbf{E}\{\Delta b_i^2\}C_{i,0} +\sum_{i=1}^{N}\sum_{j=1}^{N}\mathbf{E}\{\Delta b_i^2-\Delta b_i\Delta %b_j\}C_{i,j}\right] \text{,}
%\end{equation}
\begin{equation} \label{eq:P_coup}
{E}=\frac{V_{dd}^2}{2}\left(\sum_{i}^N\mathbf{E}\{\Delta b_i^2\}C_{i,i} +\sum_{i,j}^N\mathbf{E}\{\Delta b_i^2-\Delta b_i\Delta b_j\}C_{i,j}\right) \text{.}
\end{equation} 
%In Eq.~\ref{eq:P_coup}, the first term $\nicefrac{V^2_{dd}}{2}$ depends on the energy supply voltage $V_{dd}$. Therefore, it is a technology dependent constant. In the following ,  we describe the energy consumption normalized by this factor ($E_n$) to simplify the formulas.

Here, $C_{i,i}$ is the ground capacitance of interconnect $i$, and $C_{i,j}$ is the coupling capacitance between the interconnects $i$ and $j$. Furthermore, $\mathbf{E\{\}}$ is the expectation operator and $\Delta b_i$ represents the switching of bit $i$, which is either 1 (0 to 1 transition), 0 (no transition), or $-$1 (1 to 0 transition).
%$\Delta b_i$ is 1 for a logical 0 to logical 1 transition, -1 for a logical 1 to logical 0 transition, and 0 if the bit is stable (no transition).  
Thus, $\mathbf{E}\{\Delta b^2_i\}$ is the self switching probability of interconnect $i$.

While the energy consumption due to the ground capacitance of an interconnect $i$ is determined only by its self switching $\Delta b_i$, the energy consumption associated with a coupling capacitance $C_{i,j}$ is additionally affected by a switching on interconnect $j$ $\Delta b_j$ (correlated switching). Compared to the scenario where only interconnect $i$ toggles ($\Delta b_j=0$), the contribution of $C_{i,j}$ to the energy consumption is doubled when interconnect $j$ toggles in the opposite direction ($\Delta b_i\Delta b_j=-1$) and vanishes if it toggles in the same direction ($\Delta b_i\Delta b_j=1$).

%%%%% MAYBE SAVE FOR LATER???%%%%%%%%%%%%
% The normalized energy consumption ${P}_n$, can also  be expressed using Frobenius inner product ($\langle \rangle $) and two matrices $\mathbf{T}$ and $\mathbf{C}$: 
%\begin{equation} \label{eq:P_matr}
%{P}_n=\langle \mathbf{T}, \mathbf{C}\rangle  \text{.}
%\end{equation}
% % % %Alternative ohne function
%Here, $\mathbf{C}$ is the capacitance matrix, with capacitance $C_{i,j}$ on entry $i$,$j$.
%where $\mathbf{C}_{i,i}$ is the ground capacitance of interconnect $i$ and $\mathbf{C}_{i,j}$ is the coupling capacitance between the interconnects $i$ and $j$.
% % % % % % % % % % % % % % % %
%$\mathbf{T}$ presents the switching probabilities of the bits:
%\begin{equation}
%\mathbf{T}=\mathbf{T_{s}}\mathbf{1}_{N\times N}-\mathbf{T_{c}} \text{,}
%\end{equation}
%where $\mathbf{T_{s}}$ is a matrix with the self switching probabilities $\mathbf{E}\{\Delta b_i^2 \}$ on the diagonal entries, and zeros on the remaining entries. $\mathbf{T_{c}}$ represents the coupling probabilities with zeros on the diagonal entries and $\mathbf{E}\{\Delta b_i\Delta b_j\}$ on entry $i,j$.  $\mathbf{1_{N\times N}}$ is a matrix of ones.
In modern links, the coupling capacitances dominate over the ground capacitances~\cite{bamberg2017edgeeffects,5752410}. Therefore, recent low-power data encoding approaches aim for an increase in the correlated switching ($\mathbf{E}\{\Delta b_i\Delta b_j\}$) and a decrease in the self switching ($\mathbf{E}\{\Delta b_i^2\}$).  
However, for 3D links composed of TSVs, the logical-bit probabilities also affect the energy consumption~\cite{7961256}.
A TSV, its oxide liner and the conductive substrate form a metal-oxide-semiconductor (MOS) junction.
Thus, due to the MOS-effect, each TSV is surrounded by a depletion region, which is an insulating region within the conductive substrate.
For the typical acceptor (\textsf{p}) doped substrate, an increase of 1-bit probability ($\mathbf{E}\{b_i\}$) on a TSV enlarges the width of its surrounding depletion region. This further isolates the TSV from the conductive substrate,
resulting in up to \unit[40]{\%} lower capacitance values and consequently to reduced energy needs~\cite{7961256}.
The exact bit probability --- capacitance relation is complex and consequently not suitable for high-level models. Thus, a simple linear model is used to estimate the capacitance values $C_{i,j}$ as a function of the bit probabilities~\cite{7961256}: %Thus, the following equation is used to estimate the size of the TSV capacitances:
\begin{equation} \label{eq:cap_trad}
C_{i,j}=C_{\text{T0},i,j}+{\Delta C_{\text{T},i,j}}(\mathbf{E}\{b_i\}+\mathbf{E}\{b_j\}) \text{,}
\end{equation}
where $C_{\text{T0},i,j}$ is the capacitance value for all 1-bit probabilities equal to zero. $\Delta C_{\text{T},i,j}$ is the derivation of the capacitance value with increasing bit probability $\mathbf{E}\{b_i\}$ or $\mathbf{E}\{b_j\}$. % $\Delta C_{\text{T0},i,j}$ is obtained by extracting the capacitance values for all bit probabilities equal to one $C_{\text{T1},i,j}$, subtracting $C_{\text{T0},i,j}$ and dividing the result by two.

Summarized, the energy consumption of 3D links, normalized by the technology factor $\nicefrac{V^2_{dd}}{2}$, is estimated via:
\begin{multline} \label{eq:P_coup_3D}
  {E}_{n,3D}= \sum_{i}\mathbf{E}\{\Delta b_i^2\}\big(C_{\text{T0},i,i}+\Delta C_{\text{T},i,i}2\mathbf{E}\{b_i\}\big) \\
  +\sum_{i \neq j}\mathbf{E}\{\Delta b_i^2-\Delta b_i\Delta b_j\} 
  \big(C_{\text{T0},i,j}+\Delta C_{\text{T},i,j}(\mathbf{E}\{b_i\}+\mathbf{E}\{b_j\})\big) \text{.}
\end{multline}
For 2D links, composed of metal-wires, the capacitance quantities are independent of the pattern properties. Therefore, the energy consumption is estimated via: 
\begin{multline} \label{eq:P_coup_2D}
  {E}_{n,2D}= \sum_{i}\mathbf{E}\{\Delta b_i^2\}C_{\text{M},i,i} 
  +\sum_{i \neq j}\mathbf{E}\{\Delta b_i^2-\Delta b_i\Delta b_j\} 
  C_{\text{M},i,j}\text{.}
\end{multline}
The energy consumption can also  be expressed using Frobenius inner product ($\langle \rangle $) of two matrices, which simplifies the formulas. For 2D links we obtain: 
\begin{equation} \label{eq:P_matr}
{E}_{n,2D}=\langle \mathbf{T}, \mathbf{C}_M\rangle  \text{.}
\end{equation}
% % % %Alternative ohne function
Here, $\mathbf{C}_M$ is the capacitance matrix, with $C_{M,i,j}$ on entry $(i,j)$.
%where $\mathbf{C}_{i,i}$ is the ground capacitance of interconnect $i$ and $\mathbf{C}_{i,j}$ is the coupling capacitance between the interconnects $i$ and $j$.
% % % % % % % % % % % % % % % %
$\mathbf{T}$ presents the switching properties of the bits:
\begin{equation}
\mathbf{T}=\vec{t}_{s}\mathbf{1}_{1\times N}-\mathbf{T_{c}} \text{,}
\end{equation}
where the vector $\vec{t}_{s}$ contains the self switching probabilities $\mathbf{E}\{\Delta b_i^2 \}$. Matrix $\mathbf{T_{c}}$ includes the mean correlated switching quantities with zeros on the diagonal and $\mathbf{E}\{\Delta b_i\Delta b_j\}$ on entry $(i,j)$.  $\mathbf{1}_{1\times N}$ is a $1$$\times$$N$ matrix of ones.

For 3D links the matrix formulations is:
\begin{equation} \label{eq:P_matr_3D}
  {E}_{n,3D}=\langle \mathbf{T}, \left(\mathbf{C}_{T0}-\mathbf{\boldsymbol{\Delta}C}_{T}\circ(\vec{p}\cdot\mathbf{1}_{1\times N}+\mathbf{1}_{N\times 1}\cdot {\vec{p}}^{\, T})\right)\rangle  \text{,}
\end{equation}
where $\circ$ is the Hadamard operator and $\vec{p}$ is the bit probability vector (${p}_i=\mathbf{E}\{b_i\}$) and ${\vec{p}}^{\, T}$ its transpose.

In Summary, to estimate link energy requirements, three parameters are needed: first the capacitance matrices; second the switching matrices $\mathbf{T}$; and third the bit probabilities $\vec{p}$. While many works propose models to estimate the capacitance quantities at high abstraction levels (e.g.\ \cite{bamberg2017edgeeffects,7961256, 5613162}), the estimation of bit properties ($\mathbf{T}$ and $\vec{p}$\,) on high abstraction levels is currently possible only if virtual channels are neglected~\cite{Landman1995DBT,jafarzadeh2014data}. To consider the effect of virtual channels, cost extensive bit-level simulations are currently required. 

\section{Virtual channel-based NoCs}\label{sec:eff}
In this section we discuss the effect of virtual channels on the link energy consumption.
For this purpose, in Subsec.~\ref{sec:eff_1}, we review the concept of virtual channels.
Afterwards, in Subsec.~\ref{sec:eff_2}, we show that they have a huge impact on the link energy consumption and
heavily reduce the efficiency of existing low-power coding approaches. 
This also serves as a motivation for the present work.
\subsection{Concept of virtual channels}\label{sec:eff_1}
The main functionality of a NoC is to forward data/messages (in form of packets) from a source processing element to a destination one.
Thereby, packets may pass multiple routers (multi-hop transmission). Also, packets may compete for links if multiple of them traverse the network. %simultaneously.

Consider for example the scenario where two packets \textsf{A} and \textsf{B} compete for a link. 
Without virtual channels, one packet (e.g.\ \textsf{A}) will be granted to use the link/channel, while the second packet \textsf{B} is blocked until \textsf{A} is completely transmitted. 
This has two disadvantages: first, due to the commonly applied flit-level flow control~\cite{dallyVirtualChannelFlowControl}, the next flit of the granted packet \textsf{A} might be blocked due to a contention elsewhere in the network. In this scenario the link is idle even though the other packet \textsf{B} could make effective use of it. The second problem is that blocked packets result heavily in unequal transmission times for messages. To mitigate this degradation and to provide quality-of-service (QoS) quantities, the bandwidth of a link is divided among different packets using virtual channels. With this technique more than one buffer is associated with each input port, so that different packets are buffered simultaneously and interleaved. While the assignment of virtual channels (input buffers) is packet based, the arbitration for physical channel bandwidth is on a flit-by-flit basis~\cite{dallyVirtualChannelFlowControl}. 

Different techniques exist for this arbitration. The most common one is time multiplexing~\cite{kavaldjiev2004virtual, marescaux2002interconnection}. Hereby, the available bandwidth of the link is equally partitioned on each virtual channel (fairness).
For example, if three packets (\textsf{A} to \textsf{C}) simultaneously request the usage of one physical channel, the available link bandwidth is shared by transmitting \textsf{A$_0$B$_0$C$_0$A$_1$B$_1$C$_1$\ldots A$_m$B$_m$C$_m$} if no congestions occur in the preceding paths. Here, the indices are the flit numbers of the packets containing $m$ flits. Another common technique uses priorities~\cite{bolotin2004qnoc}.
Each virtual channel is associated with a different priority depending on the service class of the according message. The transmission of packets with a lower priority is preempted if higher priority one are using the link. This guarantees quality of service for high priority traffic at the cost of a being less fair.  

Summarized, virtual channels not only transmit packets sequentially but also multiplex simultaneous transmissions of multiple packets. The multiplexed transmission probability is slightly lower than a priority based virtual channel arbitration. 
\subsection{Effect of virtual channels on the energy consumption}\label{sec:eff_2}
As outlined in Sec.~\ref{Sec1}, the two bit-level properties that affect link power consumption are switching and bit probabilities. 
Obviously, the interleaved transmission of multiple data streams (multiplexing) - due to the use of virtual channels - has no impact on bit probabilities but on switching properties as they are determined by bit-level deltas between consecutively transmitted pattern pairs. 
 
For the sequential transmission of highly correlated data streams, as found in DSP applications, the switching properties ($\mathbf{T}$-entries) are small compared to the transmission of uncorrelated data~\cite{Landman1995DBT}. This results in relatively low energy requirements as the energy consumption is directly proportional to the switching properties. However, when the data streams are multiplexed due to the use of virtual channels, this beneficial behavior is lost as the individual messages are uncorrelated. Consequently, the link energy consumption drastically increases when virtual channels are used. This is validated by the following analysis: we investigate the mean energy consumption per transmitted byte for the transmission of two data streams containing 100,000 Gaussian distributed \unit[16]{b} flits with a relative correlation $\rho$ of 0.99 and a standard-deviation $\sigma$ of 256. Two physical link structures are considered: a 3D TSV array and a 2D metal-wire bus, both driven by commercial \unit[40]{nm} inverters. To obtain the metal wire parasitics, a commercial wire tool is used. Hereby, the metal wire width and spacing is set to \unit[0.3]{$\mu$m} and \unit[0.6]{$\mu$m}, respectively. The TSV parasitics are generated with the edge-effect-aware capacitance model from Ref.~\cite{bamberg2017edgeeffects}. For the quadratic TSV array, we consider a TSV radius and pitch of \unit[2]{$\mu$m} and \unit[8]{$\mu$m} respectively, which corresponds to the minimum TSV dimensions predicted by the International Technology Roadmap for Semiconductors (ITRS) for the time frame 2015--2018 \cite{ITRS13}. The link length is set to \unit[50]{\textmu m} (3D) and \unit[100]{\textmu m} (2D). The driver parasitics are provided by~the~vendor.

\begin{figure}[t]%		 	\centering
	\setlength\figureheight{5cm} 
	\setlength\figurewidth{0.4\textwidth}
	\centering
	% This file was created by matplotlib2tikz v0.6.15.

\begin{minipage}{.49\textwidth}

\begin{tikzpicture}
%\footnotesize
\begin{axis}[
title={\textit{a)}},
title style={at={(-0.2, -0.33)}, font = \scriptsize},
xlabel={Mux. prob.},
ylabel={Energy [\unitfrac{fJ}{B}]},
xmin=0, xmax=1,
ymin=3e-14, ymax=1.1e-13,
ytick={4e-14, 6e-14, 8e-14, 10e-14},
xtick={0, 0,2, 0.4, 0.6, 0.8, 1},
x tick label style={font = \scriptsize},
y tick label style={font = \scriptsize},
ylabel shift = 1.1em,
yticklabels={20, 30, 40, 50},
%axis on top,
at={(0\figurewidth,0\figureheight)},
width=1\figurewidth,
height=\figureheight,
%tick pos=both,
%xmajorgrids,
%ymajorgrids,
legend entries={{2D link},{3D link}},
legend cell align={left},
axis x line*=bottom,
axis y line*=left, 
ylabel style={yshift=-0.3em, font = \scriptsize},
xlabel style={yshift=0.2em, font = \scriptsize},
legend style={at={(0.1,1)}, anchor=north west,legend cell align=left,align=left,draw=white!15!black, font = \scriptsize}
]
\addplot [red, mark=*, mark size=2.5, mark options={solid,draw=black}]
table {%
	0 3.40657854202441e-14
	0.1 3.75426053055729e-14
	0.2 4.08618420511713e-14
	0.3 4.41787464880998e-14
	0.4 4.76331972390024e-14
	0.5 5.09266503475586e-14
	0.6 5.48268357563573e-14
	0.7 5.7593316012425e-14
	0.8 6.19392793528474e-14
	0.9 6.46903649861442e-14
	1 6.8357824152802e-14
};
\addplot [blue, mark=square*, mark size=2, mark options={solid,draw=black}]
table {%
	0 5.45201388094298e-14
	0.1 5.88909818628248e-14
	0.2 6.30884734037506e-14
	0.3 6.72795087493015e-14
	0.4 7.16458187697837e-14
	0.5 7.58442652058988e-14
	0.6 8.07108217188742e-14
	0.7 8.42583244170784e-14
	0.8 8.96899471000117e-14
	0.9 9.31605854375573e-14
	1 9.77630961822039e-14
};
\end{axis}

\end{tikzpicture}

\end{minipage}\hfill
\begin{minipage}{.49\textwidth}

\begin{tikzpicture}

\begin{axis}[
title={\textit{b)}},
title style={at={(-0.2, -0.33)},font = \scriptsize },
xlabel={Mux. prob.},
ylabel={Coding gain [\unit{\%}]},
xmin=0, xmax=1,
ymin=-12, ymax=16.3,
%axis on top,
at={(0\figurewidth,-0.99\figureheight)},
width=\figurewidth,
height=\figureheight,
%xmajorgrids,
%ymajorgrids,
%legend entries={{2D link},{3D link}},
legend cell align={left},
legend entries={{2D link},{3D link}},
legend cell align={left},
axis x line*=bottom,
axis y line*=left, 
x tick label style={font = \scriptsize},
y tick label style={font = \scriptsize},
ylabel style={yshift=-0.3em, font = \scriptsize},
xlabel style={yshift=0.2em, font = \scriptsize},
legend style={at={(0.1,.1)}, anchor=south west,legend cell align=left,align=left,draw=white!15!black, font = \scriptsize}
]
\addplot [red, mark=*, mark size=2.5, mark options={solid,draw=black}]
table {%
	0 14.9218058031058
	0.1 12.8478838890958
	0.2 10.8217688593667
	0.3 8.74189155070675
	0.4 6.66172586785266
	0.5 4.70155490972467
	0.6 2.68034415333352
	0.7 0.661772035425179
	0.8 -1.4036104802899
	0.9 -3.39576516279381
	1 -5.49602973603611
};
\addplot [blue, mark=square*, mark size=2, mark options={solid,draw=black}]
table {%
	0 14.6664718899538
	0.1 12.5812251390204
	0.2 10.5447213268472
	0.3 8.4572844313485
	0.4 6.36710869225543
	0.5 4.38521239387625
	0.6 2.37800846338585
	0.7 0.344107532302962
	0.8 -1.73912158763492
	0.9 -3.72883957116468
	1 -5.85301784517747
};
\end{axis}

\end{tikzpicture}
\end{minipage}
	\caption{Effect of virtual channels:
		\textit{a)} mean link energy consumption over the channel multiplexing probability for the transmission of correlated, Gaussian data streams; 
		\textit{b)} gain of classical invert-coding~\cite{stan1995bus} over the multiplexing probability for the transmission of two completely random data streams.}\label{fig:eff} 
\end{figure}
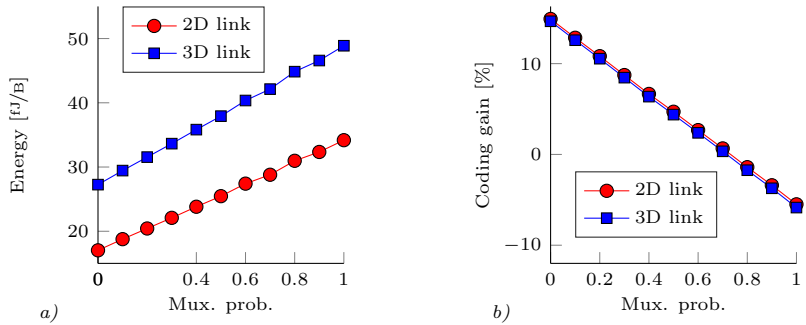

Employing bit-level simulations, we analyze the energy consumption for various data stream multiplexing probabilities (mux.\ prob.), to cover all possible scenarios in virtual channel based NoCs. Mux.\ prob.\ defines the likelihood of a change in the active virtual channel, so that the next transmitted flit belongs to another data stream than the current one.
The results are presented in Fig.~\ref{fig:eff}-a.
For no data stream multiplexing, the energy consumption of the 2D and 3D link is 17 and 28 \unit{fJ} per transmitted byte, respectively. This represents the case where no virtual channels are used. When they are used with equal priorities, mux.\ prob.\ is maximized to 1, indicating a continuous flit-by-flit multiplexing. In this scenario the energy consumption of the links is approx.\ doubled (2D: \unit[34]{fJ}; 3D: \unit[50]{fJ}) compared to the scenario where virtual channels are unused. This shows the dramatic effect of virtual channels on the link energy consumption. For scenarios where virtual channels are sometimes used, or if priorities are assigned to the channels, mux.\ prob.\ lies between 0 and 1 and the degradation is smaller.

To illustrate the effect of virtual channels on the efficiency of existing low-power codes, we analyze the transmission of two random (uniformly distributed and uncorrelated) data streams, encoded with the classical invert technique~\cite{stan1995bus}.
The results for the 2D and the 3D link, are shown in Fig.~\ref{fig:eff}-b. 
When the two data streams are transmitted without virtual channels (i.e.\ mux.\ prob.\ $= 0$), the encoding technique leads to a reduction in the 2D/3D link energy consumption by approx. \unit[15]{\%}. However, with increasing virtual channel usage (i.e.\ mux.\ prob.) the coding efficiency vanishes. Due to the added redundancy of the technique (invert-bit), the encoding approach can even increase the energy consumption by up to \unit[6]{\%}.

Thus, the high-level model presented in this work is not only required to estimate the energy consumption, but also to derive new coding techniques which do not show this degradation. 
%The previous considerations reveal the strong need for a high-level method which enables the estimation of the pattern dependent energy consumption of NoC links with virtual channels. First, for a fast estimation of the energy requirements at early design stages, and second, to enable the design and the evaluation of low-power coding techniques for 2D and 3D NoCs with virtual channels.

\section{Modeling approach}\label{sec:model}
In this section we present our high-level model to estimate the switching ($\mathbf{T})$ and bit probability ($\vec{p}$\,) characteristics of links with virtual channels, transmitting up to $n$ different data types ($D^1$ to  $D^n$). Thereby, head-flits build one data type. Thus, the amount of different transmitted message types is $n$$-$$1$.
For each individual data type, we can obtain the switching properties ($\mathbf{T^1}$ to $\mathbf{T^n}$) for a sequential transmission of the according data stream, and the bit probabilities~\cite{Landman1995DBT,AGO2011businvert}.
For our approach, not only the bit probability vectors of the data streams (${\vec{p}}^{\,1}$ to ${\vec{p}}^{\, n}$) are required, but bit probability matrices ($\mathbf{S^1}$ to $\mathbf{S^n}$), with
\begin{equation}
\mathbf{S^x}_{i,j} = \mathbf{E} \{ b^x_i \cdot b^x_j\} \text{.}
\end{equation}  
The diagonals of these matrices are equal to the bit probability vectors, since $\mathbf{E} \{ b^x_i \cdot b^x_i\}=\mathbf{E} \{ b^x_i\}$. The remaining entries of a $\mathbf{S}$-matrix are equal to the probability that both bits $i$ and $j$ of a pattern of $D^x$ are logical 1. Please note that, due to possible spatial bit-correlations (e.g.\ due to a normal distribution~\cite{Landman1995DBT}), $\mathbf{E}\{ b^x_i \cdot b^x_j\}$ is generally unequal to $\mathbf{E}\{ b^x_i\}\cdot \mathbf{E}\{b^x_j\}={{\vec{p}}^{\,x}}_i\cdot {{\vec{p}}^{\,x}}_j$.

With the data flow independent $\mathbf{S}$-matrices, we estimate the switching properties when two data streams $D^x$ and $D^y$ are multiplexed $\mathbf{T^{x\rightarrow y}}$. For this purpose, the mean self switching (${\mathbf{E}\{\Delta b_i^2\}}^{x\rightarrow y}$), as well as the mean correlated switching (${\mathbf{E}\{\Delta b_i \Delta b_j \}}^{x\rightarrow y}$ for $i \neq j$) is required. Both can be calculated~via
\begin{align} \label{eq:T_mux}
  {\mathbf{E}\{\Delta b_i \Delta b_j \}}^{x\rightarrow y} &= \mathbf{E}\{(b_i^y - b_i^x)(b_j^y - b_j^x) \} \\
  &= \mathbf{E}\{{b_i^y}b_j^y  + b_i^x{b_j^x} - b_i^y{b_j^x} - {b_i^x}b_j^y\} \nonumber \\ 
  %= \mathbf{E}\{b_i^yb_j^y\}  + \mathbf{E}\{b_i^xb_j^x\} - \mathbf{E}\{b_i^yb_j^x\} - \mathbf{E}\{b_i^xb_j^y\} \\
  &= S_{i,j}^y + S_{i,j}^x - {S_{i,i}^y}S_{j,j}^x - {S_{i,i}^x}S_{j,j}^y \nonumber \text{.}
\end{align}
For $i=j$, we obtain the self switching probabilities and for $i\neq j$ the correlated switching properties. In Eq.~\ref{eq:T_mux}, we exploit that the cross-correlation of two different data streams is zero, which results in $\mathbf{E}\{{b_i^y}b_j^x\}= \mathbf{E}\{b_i^y\}\cdot \mathbf{E}\{b_j^x\} = {S_{i,i}^y}\cdot S_{j,j}^x$.

Employing the resulting switching matrices for the scenarios of multiplexed data streams ($\mathbf{T^{x\rightarrow y}}$, with $x \neq y$), as well as the switching matrices for no multiplexing
($\mathbf{T^{x}}=\mathbf{T^{x\rightarrow x}}$), we can determine the switching for a link and a given data flow:
%This switching matrix $\mathbf{T_{link}}$ is determined via
\begin{equation} \label{eq:T_mux2}
  \mathbf{T}_{link} = \sum_{x,y}({M}_{x,y}+{M}_{x+n,y})\mathbf{T^{x\rightarrow y}} \text{,}
\end{equation}
where $\mathbf{M}$ is a $2n$$\times$$2n$ matrix which contains the information about the data flow over the link. The $(x,y)$-entry $M_{x,y}$ is equal to the probability of  transmitting a pattern of data type $y$, after transmitting a pattern of data type $x$. Thus, ${M}_{x,x}$ is equal to the probability of two subsequently transmitted patterns belonging to data type $x$ (no multiplexing). Entry $(x,x+n)$ is equal to the probability that the link holds a pattern of data type $x$ (link is idle). Therefore, entry $(x+n, y)$ is the probability of transmitting a pattern of data type $y$ after being idle, holding a value of type $x$.

Analogously, the bit probability vector for the link can be calculated by
\begin{equation} \label{eq:prob_mux}
  \vec{p}_{link} = \sum_{x,y}({M}_{x,y}+{M}_{x,y+n}+{M}_{x+n,y}){\vec{p}}^{\, y} \text{.}
\end{equation}
Finally, by substituting $\mathbf{T}_{link}$ and $\vec{p}_{link}$ into Eq.~\ref{eq:P_matr} (for a 2D link) or Eq.~\ref{eq:P_matr_3D} (for a 3D link) we can estimate the energy consumption of links in the presence of virtual channels.

\section{Simulator}\label{sec:simulator}

In this section, we describe our simulator, which is capable of generating the data flow matrices $\mathbf{M}$. First, we introduce its architecture. Second, we demonstrate the configuration options of the simulator. Third, we explain the generation of data flow matrices and other evaluation metrics. The simulator and its source code are publicly available on Github at \url{https://github.com/jmjos/ratatoskr}. Originally, it targets NoCs for 3D SoCs with technology heterogeneity, but can be used to model traditional 2D and homogeneous 3D NoCs, as well.

\subsection{Architecture}

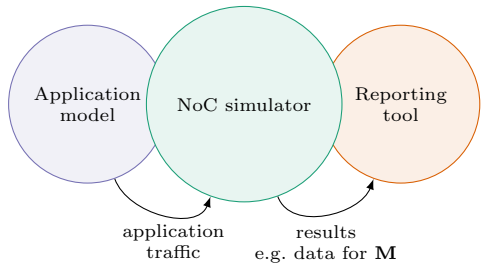
\begin{figure}
	\centering
	\begin{tikzpicture}
	\node[circle, black, draw = col3, fill = col3!10, desc, align = center, inner sep=0pt, minimum size = 2.1cm] (app) {Application\\ model};
		\node[circle, black, draw = col2, fill = col2!10, desc, align = center, inner sep=0pt, minimum size = 2.1cm, right= 2.05cm of app] (report) {Reporting\\ tool};
	\node[circle, black, draw = col1, fill = col1!10, desc, inner sep=0pt, minimum size = 2.6cm] at ($(app)!0.5!(report)$) (nocsim) {NoC simulator};
	\draw[-latex] (app) to [out=-70,in=-110] (nocsim) node [yshift = -7pt, xshift = -14pt, desc, below, align = center, anchor = north] {application\\ traffic};
	\draw[-latex] (nocsim)   to [out=-70,in=-110] (report)node [yshift = -14pt, xshift = -18pt, desc, below, align = center, anchor = north] {results\\e.g.\ data for $\mathbf{M}$};

\end{tikzpicture}
	\caption{Components of the simulator}
	\label{fig:components}
\end{figure}

The simulator consists of three individual parts as shown in Fig.\ \ref{fig:components}: The \emph{NoC simulator} is the core of the software; it simulates the NoC by means of hardware models for its components. The \emph{application model} provides an implementation, which injects synthetic or real-world based traffic into the network. The \emph{reporting tool} offers functions for evaluation of the simulation results. Neither application model nor reporting tool are topics of this very publication, and we kindly refer to the given references for details. We only provide a short description of the parts, while the NoC simulator is described in more detail in the next paragraph. The three parts are implemented in C++11 using the class library SystemC 2.3.1a \cite{SystemCStandard}, which provides the simulation kernel. In version 1.1.8, the simulator has approximately 7,700 lines of code. This is a similar code size as the competitors (Noxim: 8,600 lines of code). Please note that BookSim 2.0 provides its own kernel and thus is naturally larger with 25,000 lines of code.

The application model implements colored, statistical Petri nets with retention time on places as published in \cite{Joseph.2018}. This allows to model a very wide range of applications, reassessing both real world based traffic streams and traditional, synthetic ones. The reporting tool, as proposed in \cite{Joseph.2016c}, generates textual and graphical reports for each simulation run of the simulator. Further, it offers a MySQL database to track events in the NoC simulator such as \emph{flit send} or \emph{routing calculation}. It also generates the data flow matrices as explained later on. 

\begin{figure}
	\centering
	\pgfdeclarelayer{background}
\pgfdeclarelayer{foreground}
\pgfsetlayers{background,main,foreground}   %% some additional layers for demo
\begin{tikzpicture}[scale = .60,
block/.style={
	draw,
	fill=white,
	rectangle, rounded corners,
	minimum width={width("Processing element")+2pt},
	minimum height={height("Processing element")+2pt},
	desc}, 
titleblock/.style={
	desc}, 
level/.style={desc}
]

\node[titleblock](model){\bfseries };

\node[block, below right = 0.7cm and 0.3cm of model] (pe) {Processing element};
\node[block, below=0.5cm of pe] (ni) {Network interface};
\node[block, below=0.5cm of ni] (router) {Router};
\node[block, below=0.5cm of router] (link) {Link};
\draw[latex-latex] (pe) -- (ni)node[midway, right, desc] {\scriptsize Packet};
\draw[latex-latex] (ni) -- (router)node[midway, right, desc] {\scriptsize Flit};
\draw[latex-latex] (router) -- (link)node[midway, right, desc] {\scriptsize Flit};

\node[block, below left =0.7cm and .3 cm of model] (app) {Task};

\node[titleblock, fill = white, above=0.2cm of pe] (hw) {\emph{NoC model}};
\node[titleblock, fill = white,  above = 0.2cm of app] (appmodel) {\emph{Application model}};

%links
%\draw (model) -| (hw.north);
%\draw (model) -| (appmodel.north);
%\draw (hw) -- (pe);
%\draw (appmodel) -- (app);
\draw[latex-latex](app) -- (pe) node[midway, desc, rotate = -90, anchor = west]{\scriptsize TLM interconnect};

%levels of abstraction 
\node [level, below = 6pt of link.south, yshift = 0, rotate =0, anchor = north, align=center] (cycleAccurate) {\bfseries cycle accurate};%\\ \bfseries level};
\node [level, anchor = north, align=center] (tlm) at (app.south|-cycleAccurate.north) {\bfseries transaction model};%\\ \bfseries model};

\begin{pgfonlayer}{background}
\filldraw [line width=4mm,col1!10]
(pe.north -| pe.west) rectangle (cycleAccurate.south -| pe.east);
\filldraw [line width=4mm,col3!10]
(app.north -| app.west) rectangle (tlm.south -| app.east);
\end{pgfonlayer}

\end{tikzpicture}
	\caption{Modules of NoC and application simulator.}
	\label{fig:architecture}
\end{figure}
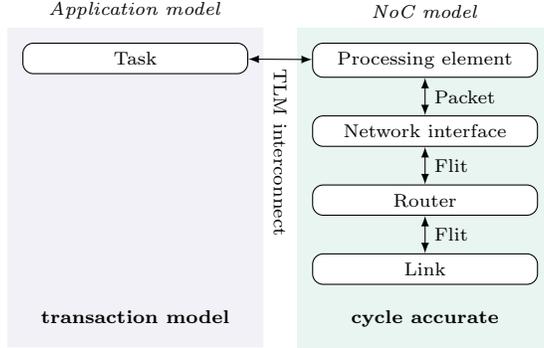

The NoC simulator implements a hardware model, consisting of PEs, NIs, routers and links. As a distinguishing factor to the competitors, the model is well-defined, as published in \cite{Joseph.2018}. The architecture, i.e.\ structure of components and their communication, is shown in Fig.~\ref{fig:architecture}. As already explained, the application model injects traffic into the network via PEs. Since the application is modeled on transaction level (left-hand side of the figure) and the hardware is modeled cycle-accurate (right-hand side), both parts are connected by a transaction level model (TLM) interconnect. It handles mapping of places in the application's Petri net to PEs and vice versa. The \emph{PE} is an abstract representation of processing cores, sensors or hardware accelerators connected to the NoC. PEs support virtual channels and are implemented in the class \texttt{ProcessingElementVC}. PEs process packets from the network via a receive-function. In the PEs' execute-function, events are triggered in the application model for the application dynamic behavior. The simulator connects PEs with routers at the same location via a single NI. NIs serialize data from packets into flits using the network link bit width, and vice versa. The NIs' implementation in the class \texttt{NetworkInterfaceVC} consists of one function per direction to serialize and deserialize data. Flits are sent in the network via routers and links. The router model is implemented in the class \texttt{RouterVC}. The router model is a standard input-buffered router \cite{Dally.2004}, which in addition, supports non-purely synchronous communication. Routers have two main functions: Flits are stored in the correct buffers in a receive-function; it also handles flow control to upstream routers. In the router's main thread, flits are processed by calculating a route, arbitrating virtual channels and sending flits. The router model has three stages (routing calculation, virtual channel allocation, sending) for head flits. Subsequent flits will be transmitted immediately if downstream buffer space is available. Many parameters of the router, such as buffer depth, number of virtual channels, routing function and even the network topology (cf.\ \cite{Joseph.2018}) can be configured during runtime using XML files, which avoids recompiling the simulator. As one exemplary excerpt, the definition of a router and a PE model is shown Listing \ref{lst:xml:nodeTypes}. The router supports virtual channels, uses deterministic dimension ordered routing, round robin selection\footnote{Actually, the selection is not relevant for deterministic routing algorithms, which only return a single path.}, a fair arbiter as introduced in Sec.~\ref{sec:eff} and a clock speed of \unit[1]{GHz}. The PEs have fewer options; they support virtual channels and are clocked at \unit[500]{MHz} in this example. Finally, links connect routers unidirectionally, and their model is implemented in the class \texttt{Link}. Links issue data transmission clock-wise to the reporting tool for the generation of data flow matrices $\mathbf{M}$. We verified the correctness of these matrices by comparison with raw data. It is noteworthy, that the link itself does not require knowledge about link bit width, since this information is already encapsulated into the flit generation in the NIs.

\begin{figure}
\begin{lstlisting}[style=myxml, caption = {Configuring node types via XML.}, label=lst:xml:nodeTypes]
<nodeTypes>
	<nodeType id="0">
		<model value="RouterVC"/>
		<routing value="XYZ"/>
		<selection value="RoundRobin"/>
		<arbitration value="fair"/>
		<clockDelay value="1"/>
	</nodeType>
	<nodeType id="1">
		<model value="ProcessingElementVC"/>
		<clockDelay value="2"/>
	</nodeType>
</nodeTypes>
\end{lstlisting}
\end{figure}

In general, the proposed architecture is similar to competing software. Important distinguishing features are as follows: The proposed simulator is the only one with a link model to calculate the dynamic energy consumption of virtual channel-based links, as proposed as part of this publication. The number of parameters, which can be set during runtime via XML files is high in comparison and allows for very flexible modeling of many architectures. The application model, as published in \cite{Joseph.2018}, is comprehensive and allows for generating real world data traffic, which is not possible with the existing solutions: These only provide synthetic traffic patterns, which are on a different level of abstraction. Finally, the reporting tool enables more flexible reports with adjustable level of abstraction than the competitors. This is exemplified in detail in the next section. To summarize, the proposed simulator offers more diverse features than the competing software and has an energy model of higher accuracy.

\subsection{Generation of the data flow matrices \textbf{M}}

\begin{figure}
	\centering
	%\documentclass[border=2mm]{standalone}
%\usepackage{pgfplots}
%\pgfplotsset{compat=1.8}
%\usepackage{amssymb}
%\usepackage{units}
%\usetikzlibrary{decorations.pathreplacing, decorations, positioning, calc, patterns, arrows, math, shadings}
%\input{defs.tex}
%
%\begin{document}
	
%#1router pos x
%#2 router pos y
%#3 layer
%#4 size pe x
%#5 size pe y
%#6 color network
%#7 color pe
%#8 circle size
\newcommand{\drawRouterLocalPE}[8]{
	\draw[#6](#1,#3,#2) -- (#1+.3,#3,#2+.3);
	\draw[draw = #6, fill=#6](#1,#3,#2) circle (#8pt);
	\draw[#7, fill = #7!50, opacity= 0.8](#1+.3, #3, #2+.3) --(#1+.3+#4, #3, #2+.3) --(#1+.3+#4, #3, #2+.3+#5) --(#1+.3, #3, #2+.3+#5) -- cycle;
}

\begin{tikzpicture}[scale = .5,
cube/.style={thick, black, col1, fill = col3, fill opacity = 0.2},
axis/.style={-latex,col2, thick}, digital/.style={thick, col1}, digitalLines/.style = {col1}]

\tikzmath{\l1 = 0;};
\tikzmath{\l2 = -3.5;};
\tikzmath{\l3 = 7;};

%noc layer
\draw[col1!50,  fill = col1!10] (0.8,\l2, 0.8) -- (0.8,\l2,9.2) --(9.2,\l2,9.2) -- (9.2,\l2,0.8) -- cycle;
\draw (3,\l2, 3) -- (7,\l2, 3) --(7,\l2, 7) -- (3,\l2, 7) -- cycle;
\draw[fill = gray] (3,\l2, 3) circle (.2);
\draw[fill = gray] (7,\l2, 3) circle (.2);
\draw[fill = gray] (7,\l2,7) circle (.2);
\draw[fill = gray] (3,\l2, 7) circle (.2);

%first path
\draw[-latex, col2!90, rounded corners, thick] (7.5-0.05,\l1, 4.5) -- (7-0.05,\l2+.15, 3) -- (3,\l2+.15, 3)  -- (3,\l1-.2, 3);
\draw[-latex, red, rounded corners, thick] (7.5+0.05,\l1, 4.5) -- (7+0.05,\l2-.15, 3) -- (3,\l2-.15, 3) -- (3,\l2-.15, 7) -- (3,\l1-.2, 7);

%application layer
\draw[col3!50] (0.8,\l1, 0.8) -- (0.8,\l1,9.2) --(9.2,\l1,9.2) -- (9.2,\l1,0.8) -- cycle;
\draw[fill = col3!50, opacity= 0.1] (0.8,\l1, 0.8) -- (0.8,\l1,9.2) --(9.2,\l1,9.2) -- (9.2,\l1,0.8) -- cycle;
\draw[fill = white] (3,\l1, 3) circle (.2);
\draw[fill = white] (3,\l1, 7) circle (.2);
\draw[fill = white] (7.5,\l1, 4.5) circle (.2);
\draw[-latex] (7.5-0.3,\l1, 4.5-0.2) -- (3+0.2,\l1, 3+0.05);
\draw[-latex] (7.5-0.15,\l1, 4.5+0.2) -- (3+0.22,\l1, 7);

%labels
\node[desc, anchor = west, align = left] at (9.5,\l1, 5) (app) {Application};
\node[desc, anchor = west, align = left] at (9.5,\l2, 5) (app) {NoC};
\node[desc, anchor = south, align = center] at (5,\l2, 3) (app) {$\mathbf{M}_1\!\propto\!\textcolor{col2}{\sigma_1}, \textcolor{red}{\sigma_2}$};
\node[desc, anchor = west, align = left] at (3.2,\l2, 5.5) (app) {$\mathbf{M}_2\!\propto\!\textcolor{red}{\sigma_2}$};
\node[desc, align = right, anchor = east, red] at  (3,\l1/2+\l2/2, 7) {$\sigma_2$};
\node[desc, align = right, anchor = east, col2] at  (3,\l1/2+\l2/2, 3) {$\sigma_1$};

\end{tikzpicture}

%\end{document}
	\caption{Interplay of application model and link matrices.}
	\label{fig:links}
\end{figure}
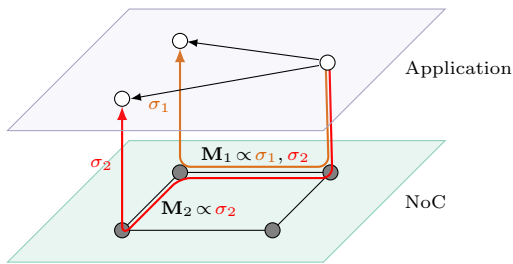

The $\mathbf{M}$-entries depend on the application scenario and the NoC architecture (virtual channel count/arbitration, routing, etc.). A typical scenario for the generation of data flow matrices is shown in Fig.~\ref{fig:links}. In the upper part of the figure, an application is shown. It consists of a sender and two receivers; the data for each receiver have different pattern types, i.e.\ different switching ($\mathbf{T}$) and bit probabilities ($\vec{p}$). In the lower layer, a simple $2\!\times\!2$ NoC is shown with mesh topology and dimension order routing. For the sake of simplicity, only routers and links are shown; NIs and PEs are at the same position of routers. The sender is mapped to the upper right PE. The senders are mapped to two PEs on the left-hand side of the NoC.\footnote{Actually, to comply with the model \cite{Joseph.2018}, two different places are required to send two different pattern types, which are mapped to the same PE. For the sake of simplicity, we depict a single place.}  In the application model, as proposed in \cite{Joseph.2018}, $n$ different pattern types are denoted by colors using the set $\Sigma =\{\sigma_1, \sigma_2, \dots, \sigma_n\}$. In this example, only $\sigma_1$ and $\sigma_2$ are used. The data flow of both pattern types is shown in the figure in red and orange. The data flow matrix $\mathbf{M}_1$ for the upper link is influenced by both pattern types; the data flow matrix $\mathbf{M}_2$ is only influenced by type $\sigma_2$, since the data of the first flow do not traverse this link. If the two pattern types are in two different virtual channels and are transmitted simultaneously, there will be switching activity between the two types in the data flow matrix for this link. This example demonstrates the dependence of data flow matrices from NoC and application.

This method is superior to saving a whole protocol of the transmitted flit types (which we used to verify the implementation of the link model), since this would require memory which linearly increases with the amount of simulated clock cycles, i.e.\ the trace of the link has a memory complexity of $O(t)$, with $t$ as simulation time. The effort to save the data flow matrices, however, is constant because the matrices are of size $2n$$\times$$2n$ for $n-1$ transmitted data streams and do not increase their size with the simulation time, i.e.\ the generation of the matrices has a memory complexity of $O(n^2)$. Naturally, the execution time complexity is identical for both methods, since the whole simulation must be executed ($O(t)$). 

The reporting tool saves the matrix's contents and reports them both in human-readable textual form and as a \texttt{.csv} file for further processing. Please note that the data flow matrices allow for the calculation of further statistics, beside the proposed energy consumption. For instance, the average idle/usage time of every link, and thus also router port, can be easily extracted. Therefore, the data flow matrices provide an innovative feature with a truly additional value in comparison to the existing NoC simulators.

\section{Simulation results}\label{sec:sim_res}
\subsection{Model accuracy}
In this section we investigate the accuracy of our approach. For this purpose, we analyze the transmission of 2--5 data types/streams for different mean multiplexing probabilities with Python. 
Furthermore, each simulation is executed 1,000 times for a flit width of \unit[16]{b}, and 1,000 times for a flit width of \unit[32]{b}. To cover the large space of data type combinations as well as possible, in each run the synthetically generated data streams vary randomly. The pattern distribution of each single data stream is either uniform, normal (Gaussian), or log-normal. For the last two distributions, the standard deviation of the patterns is in the range from $2^{\nicefrac{N}{10}}$ to $2^{N-1}$ and the relative pattern correlation is in the range from $0$ to $1$.
Compared are the switching properties ($\mathbf{T}$), estimated with our proposed high-level model, and the exact switching properties determined by means of bit-level simulations, for the transmission of 10,000 flits. Reported are the overall root-mean-square-errors (\textit{RMSE}) as well as the maximum-absolute-errors (\textit{MAE}). 
%Since, the switching properties are independent of the link type (2D or 3D),
%this particular simulation setup enables to quantify the general accuracy of our modeling approach for all possible data flow scenarios.
\begin{figure}[t]%		 	\centering
	\centering
	% !TeX spellcheck = en_US
% This file was created by matlab2tikz.
%
%The latest updates can be retrieved from
%  http://www.mathworks.com/matlabcentral/fileexchange/22022-matlab2tikz-matlab2tikz
%where you can also make suggestions and rate matlab2tikz.
%
%POSSIBLE PATTERN ARE
%horizontal lines
%vertical lines
%north east lines
%north west lines
%grid
%crosshatch
%dots
%crosshatch dots
%fivepointed stars
%sixpointed stars
%
\definecolor{mycolor1}{rgb}{0,0,0}%
\definecolor{mycolor5}{rgb}{0.5,0,0}%
\definecolor{mycolor2}{rgb}{1,0,0}%
\definecolor{mycolor4}{rgb}{1,0.3,0.3}%
\definecolor{mycolor3}{rgb}{1,1,1}%

\setlength\figureheight{2.5cm} 
\setlength\figurewidth{0.35\textwidth}

\begin{tikzpicture}
%\footnotesize
\begin{axis}[%
scaled ticks=false,
width=\figurewidth,
height=\figureheight,
at={(0\figurewidth,0\figureheight)},
scale only axis,
log origin=infty,
xmin=1.5,
legend columns=2,
xmax=5.5,
ymin = 0,
ymax = 0.008,
xtick={2, 3, 4, 5},
xlabel={Data stream count},
ytick={0, 0.004, 0.008},
yticklabels={ 0, 0.4,0.8},
ylabel={\textit{RMSE}  [\unit{pp.}]},
ylabel style={yshift=-0.3em, font = \scriptsize},
xlabel style={yshift=0.2em, font = \scriptsize},
axis background/.style={fill=white},
title={\textit{a)}},
title style={at={(-0.14, -0.44)}, font = \scriptsize},
x tick label style={font = \scriptsize},
y tick label style={font = \scriptsize},
axis x line*=bottom,
axis y line*=left,
legend style={at={(1.2,1.1)},anchor=south, column sep= 5pt,legend cell align=left,align=left,draw=white!15!black, font = \scriptsize}
]
\addplot[ybar,bar width=0.15,bar shift=-0.225,area legend ,postaction={pattern=horizontal lines, pattern color=black}] plot table[row sep=crcr] {%
2	0.006371187860749448 \\ 
3	0.0062419918348870368\\
4	0.0068363965835142844\\
5	0.0064118575762775959\\
};
\addlegendentry{Mux. prob. $=$ 0.1};

\addplot[ybar,bar width=0.15,bar shift=-0.075,area legend] plot table[row sep=crcr] {%
2   0.0065493218977126268 \\
3   0.007026968617141436  \\
4   0.0064485180724252011 \\
5   0.0063484610058446959 \\
};
\addlegendentry{Mux. prob. $=$ 0.4};

\addplot[ybar,bar width=0.15,bar shift=0.075,area legend ,postaction={pattern=dots, pattern color=black}] plot table[row sep=crcr] {%
2  0.0066481552923710614 \\
3  0.0065327700908554265 \\
4  0.0064422242906961256 \\
5  0.0070465614308143788 \\
};
\addlegendentry{Mux. prob. $=$ 0.7};

\addplot[ybar,bar width=0.15,bar shift=0.225,area legend ,postaction={pattern=north east lines, pattern color=black}] plot table[row sep=crcr] {%
2  0.0066552117167083063 \\
3  0.0069960228733102699 \\
4  0.006841786277234258  \\
5  0.0072519680804371345 \\
};
\addlegendentry{Mux. prob. $=$ 1};

\end{axis}
\begin{axis}[%
scaled ticks=false,
width=\figurewidth,
title={\textit{b)}},
title style={at={(-0.14, -0.44)}, font = \scriptsize},
height=\figureheight,
at={(1\figurewidth+1.5cm,0\figureheight)},
scale only axis,
log origin=infty,
xmin=1.5,
xmax=5.5,
ymin = 0,
xtick={2, 3, 4, 5},
xlabel={Data stream count},
ytick={0, 0.01, 0.02, 0.03},
yticklabels={ 0, 1.0, 2.0, 3.0},
ylabel={\textit{MAE}  [\unit{pp.}]},
axis background/.style={fill=white},
ylabel style={yshift=-0.3em, font = \scriptsize},
xlabel style={yshift=0.2em, font = \scriptsize},
x tick label style={font = \scriptsize},
y tick label style={font = \scriptsize},
axis x line*=bottom,
axis y line*=left,
legend style={at={(0.7,0.9)},anchor=south west,legend cell align=left,align=left,draw=white!15!black, font=\footnotesize}
]
\addplot[ybar,bar width=0.15,bar shift=-0.225,area legend ,postaction={pattern=horizontal lines, pattern color=black}] plot table[row sep=crcr] {%
2  0.022824451689569088\\
3  0.025446998628262833\\
4  0.022040064371287023\\
5  0.022573364118681882\\
};

\addplot[ybar,bar width=0.15,bar shift=-0.075,area legend] plot table[row sep=crcr] {%
2  0.022881382181018092 \\
3  0.024890290916691649 \\
4  0.02354385355155519 \\
5  0.027181276282548128 \\
};

\addplot[ybar,bar width=0.15,bar shift=0.075,area legend ,postaction={pattern=dots, pattern color=black}] plot table[row sep=crcr] {%
2  0.024412399765476597 \\
3  0.024708242768696841 \\
4  0.02414273126812688 \\
5  0.027048501365476598 \\
};

\addplot[ybar,bar width=0.15,bar shift=0.225,area legend ,postaction={pattern=north east lines, pattern color=black}] plot table[row sep=crcr] {%
2  0.024774734856485625 \\
3  0.027362035295529612 \\
4  0.027138693674867544 \\
5  0.024518005252425256 \\
};

\end{axis}

\end{tikzpicture}%
	\caption{Accuracy of our proposed high-level model compared to bit-level simulations: \textit{a)} root mean square error (RMSE); \textit{b)} maximum absolute error. Both quantities are normalized and given in percentage points (\unit{pp}).}\label{fig:sim_error} 
\end{figure}
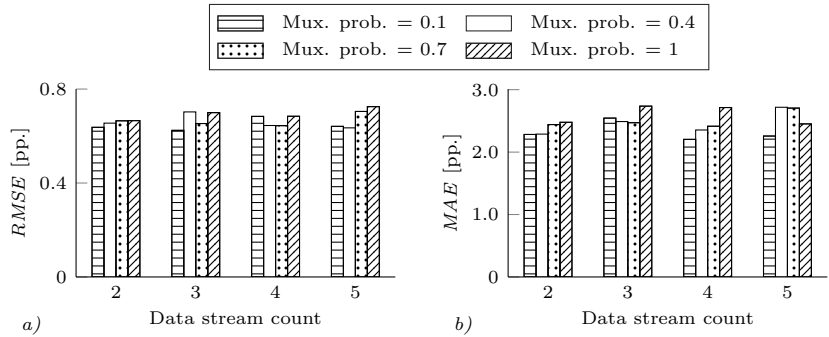

The results are presented in Fig.~\ref{fig:sim_error}. The results show that our approach enables an extremely accurate estimation of the switching characteristics, independent of the multiplexing probabilities, the number of multiplexed data streams, or the flit width. For all analyzed scenarios, the \textit{RMSE} of our estimates is in the range of \unit[0.6--0.8]{percentage points (pp)}. The maximum error (\textit{MAE}) for all 5,120,000 estimated switching properties does not exceed \unit[2.8]{pp}.
%In contrast if the effect of multiplexing is neglected, the error in the estimation of the switching characteristics, and consequently the error is the estimation of power quantities, can bigger than \unit[50]{pp} with $RMSE$ values of up to pver \unit[25]{\%}. 
Although our model has a close to perfect match with bit-level simulations, it requires more than 2,000 times lower execution time on an \textsf{Intel} i5-4690 machine, with \unit[16]{GB} of RAM, running Linux kernel 3.16. This speed up will even increase with an increasing pattern and/or link count. Thus, our model enables to precisely predict the energy consumption of a full virtual channel based NoC, containing multiple processing elements, within a tolerable time. 

Furthermore, the experiment proves that, as expected, the switching activities linearly increase with an increasing multiplexing probability (see also Fig.~\ref{fig:eff}), while the number of multiplexed data streams does not affect the switching properties. Since switching is only determined by direct consecutive pattern pairs, a multiplexing of two data streams  leads, on average, to the same energy consumption as a multiplexing of three or more data streams. Thus, without loss of generality, in the remainder of this section we restrict our analysis to scenarios where only two data types are multiplexed.

\subsection{Low-power coding}\label{sec:sim_coding}
As outlined in Sec.~\ref{sec:eff},
there is a strong need for coding techniques which reduce the energy consumption of links with virtual channels. 
The modeling technique presented in this work allows for a fast estimation of the efficiency of such low-power coding techniques. 
With a single simulation, the coding independent data flow matrices $\mathbf{M}$ are determined once. Afterwards, using Eq.~\ref{eq:P_matr}--\ref{eq:prob_mux}, the actual link energy requirements can be determined for arbitrary data streams, and thus different applied data encoding techniques. Furthermore, the high-level model presented in this work
enables the design of new coding techniques which consider the effect of virtual channels. This approach is investigated in this subsection.  

For this purpose, we consider the simultaneous transmission of \unit[2]{MB} of data from two different sources over \unit[16]{b} wide 2D and 3D links. For the physical media (2D and 3D links), we use the same structures as in Sec.~\ref{sec:eff}. Investigated is the dissipated energy per effectively transmitted byte over the mean multiplexing probability (mean$({M}_{1,2}, {M}_{2,1})$) to take possible bit-overheads of encoding techniques into account. 
For the data streams, we consider uniformly distributed patterns where the eight MSBs show a strong temporal correlation ($\rho = 0.99$), and completely random (uncorrelated) patterns.  For the uncorrelated, data we analyze the invert-coding and for the correlated data a correlator-coding~\cite{5876688}.
The second approach correlates (bit-wise \textsc{XOR}) every data word with the previous data word of the stream. Therefore, in the analyzed example, the high MSB correlation in combination with the inverting drivers leads to code word MSBs nearly stable on logical 1~\cite{garcia2010practical}.
The energy quantities are determined twice by means of Eq.~\ref{eq:P_matr}--\ref{eq:P_matr_3D}: once using the high-level model presented in this work to estimate $\mathbf{T}$ \& $\vec{p}$\,; and once using the exact $\mathbf{T}$ \& $\vec{p}$ obtained by bit-level~simulations. 
\begin{figure}%[t]%		 	\centering
	\setlength\figureheight{6cm} 
	\setlength\figurewidth{0.5\textwidth}
	\centering
	\input{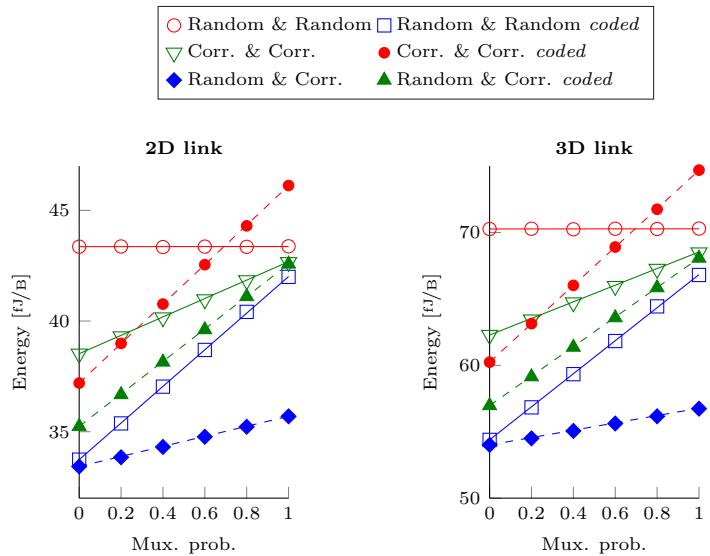}
	\caption{Effect of low-power coding techniques on the 2D/3D link energy consumption for the transmission of correlated and fully random data in the presence of virtual channels. Marks show results for exact bit-level simulations, corresponding lines are estimates of our proposed high-level model.}\label{fig:coding} 
\end{figure}

The resulting energy quantities are illustrated in Fig.~\ref{fig:coding}. The markers indicate the energy quantities obtained for the exact bit properties which are in perfect accordance with the energy quantities obtained with our high-level model (lines). Thus, our model allows for a fast and precise estimation of coding efficiencies. For example, the model precisely predicts the decreasing efficiency of the invert-coding with an increasing multiplexing probability as well as the increasing efficiency of the correlator-coding for the transmission of two correlated data streams. Thus, it allows to identify that for a high virtual channel usage a different coding technique than the invert-coding approach is required, while the correlator-coding performs well for this data flow scenario. Without our high-level model, it is not trivial for a designer to explain this observation, which complicates the design of new coding techniques for links with virtual channels. Our proposed model provides a clear answer. Invert-coding does only affect the switching probabilities for the sequential (non-multiplexed) transmission of data streams. However, it does not affect the bit probability matrices $\mathbf{S}$. Thus, according to our model, it does not decrease the energy consumption per clock cycle for a continuous data type multiplexing, and due to its induced overhead it even increases the energy consumption per effectively transmitted byte. In contrast, the correlator-coding leads to MSBs nearly stable on logical 1. This increases the $\mathbf{S}$-values, which reduces the $\mathbf{T}^{x\rightarrow y}$ values as well as the TSV capacitance quantities via the MOS-effect. 

Summarized, our model reveals an important message for the design of low-power techniques: in order to obtain the most efficient coding approach for links with virtual channels, the technique must not only affect the switching activities of the single data streams  $\mathbf{E}\{\Delta b_i\Delta b_j\}$, but also the bit probabilities~$\mathbf{E}\{b_i\cdot b_j\}$.
\section{Image processing case study}\label{sec:cs}

This section presents a case study with a common use-case of the model presented in this work. We consider a heterogeneous 3D Vision SoC with a NoC interconnect architecture. The NoC architecture has a flit width of \unit[16]{b} with 1 head- and 31 body-flits (payload) per packet, supports up to four virtual channels per port, and has an input buffer depth of 4. Furthermore, for comparison, we also consider the same architecture without virtual channels.

The full 3D SoC consists of: one mixed signal (MS) layer which contains six CMOS image sensors (S1--S6) at the top; one underlying memory (MEM) layer; and at the bottom one digital layer containing the actual processors. In this case study we want to analyze, and optimize, the transmission of raw gray-scale image pixels (two per flit) from the sensors to the memory.
Since the analyzed NoC uses a XYZ-routing~\cite{de2006networks}, while images are always read from the memory, we can analyze the traffic from the memory to the sensors without considering the traffic between the cores and the memory. 
 
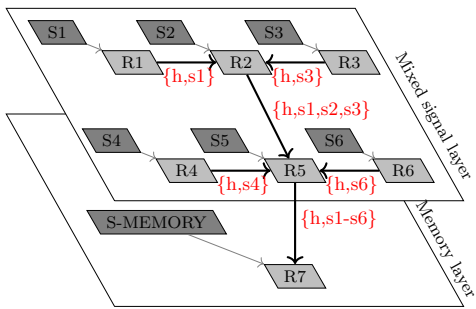
\begin{figure}	 	
	\centering
	\scalebox{0.8}{
	% This file was created by matlab2tikz.
%
%The latest updates can be retrieved from
%  http://www.mathworks.com/matlabcentral/fileexchange/22022-matlab2tikz-matlab2tikz
%where you can also make suggestions and rate matlab2tikz.
%
%REQURIED IN THE PREAMPLE
\pgfdeclarelayer{bg}    % declare background layer
\pgfsetlayers{bg,main} %set layer order
\begin{tikzpicture}
\footnotesize

%\tikzstyle{every node}=[trapezium, draw, minimum width=3cm,
%trapezium left angle=120, trapezium right angle=60] %

\draw[fill=white] (-0.8, -0.5) -- ( -2.6, 2.7) -- ( 3.2, 2.7) -- (5, -0.5) --(-0.8, -0.5) ;
\begin{pgfonlayer}{bg}
\draw[yshift=-50] (-0.8, -0.5) -- ( -2.6, 2.7) -- ( 3.2, 2.7) -- (5, -0.5) --(-0.8, -0.5) ;
\end{pgfonlayer}

\node[fill=lightgray,trapezium, draw,  trapezium left angle=120, trapezium right angle=60](R1)
    at (-0.5,1.8) {R1};
\node[fill=lightgray,trapezium, draw,  trapezium left angle=120, trapezium right angle=60](R2)
    at (1.3,1.8) {R2};
\node[fill=lightgray,trapezium, draw,  trapezium left angle=120, trapezium right angle=60](R3)
    at (3.1,1.8) {R3};

\node[rotate=299] at (4.5, 0.9) {Mixed signal layer};
\node[yshift=-50,rotate=299] at (4.7, 0.45) {Memory layer};

\node[fill=lightgray,trapezium, draw,  trapezium left angle=120, trapezium right angle=60](R4)
    at (0.4,0) {R4};
\node[fill=lightgray,trapezium, draw,  trapezium left angle=120, trapezium right angle=60](R5)
    at (2.2,0) {R5};

\node[fill=lightgray,trapezium, draw,  trapezium left angle=120, trapezium right angle=60](R6)
    at (4,0) {R6};
\node[yshift=-50,fill=lightgray,trapezium, draw,  trapezium left angle=120, trapezium right angle=60](R7)
    at (2.2,0) {R7};

\draw[->, line width=1pt] (R1)-- node[below, red,yshift=1,xshift=1] {\small \{h,s1\}}(R2);
\draw[->, line width=1pt] (R3)--  node[below, red,yshift=1,xshift=1] {\small \{h,s3\}}(R2);
\draw[->, line width=1pt] (R2)--  node[right, red,yshift=3,xshift=-1.5] {\small \{h,s1,s2,s3\}}(R5);
\draw[->, line width=1pt] (R4)-- node[below, red,yshift=1,xshift=1] {\small \{h,s4\}}(R5);
\draw[->, line width=1pt] (R6)-- node[below, red,yshift=1,xshift=1] {\small \{h,s6\}}(R5);
\draw[->, line width=1pt] (R5)--node[right, red,yshift=2,xshift=-1] {\small \{h,s1-s6\}}(R7);

\node[trapezium, draw, fill=gray,  trapezium left angle=120, trapezium right angle=60](S1)
    at (-1.75,2.3) {S1};
\node[trapezium, draw, fill=gray,  trapezium left angle=120, trapezium right angle=60](S2)
    at (0.05,2.3) {S2};
\node[trapezium, draw, fill=gray,  trapezium left angle=120, trapezium right angle=60](S3)
    at (1.85,2.3) {S3};
\node[trapezium, draw, fill=gray,  trapezium left angle=120, trapezium right angle=60](S4)
    at (-0.85,0.5) {S4};
\node[trapezium, draw, fill=gray,  trapezium left angle=120, trapezium right angle=60](S5)
    at (0.95,0.5) {S5};
\node[trapezium, draw, fill=gray,  trapezium left angle=120, trapezium right angle=60](S6)
    at (2.85,0.5) {S6};
\node[yshift=-50,trapezium, draw, fill=gray,  trapezium left angle=120, trapezium right angle=60](S7)
    at (-0.1,0.9) {S-MEMORY};
\draw[->, gray] (S1)--(R1);
\draw[->, gray] (S2)--(R2);
\draw[->, gray] (S3)--(R3);
\draw[->, gray] (S4)--(R4);
\draw[->, gray] (S5)--(R5);
\draw[->, gray] (S6)--(R6);
\draw[->, gray] (S7)--(R7);

%%%Hex 7
%line 1

\end{tikzpicture}%
}
	\caption{Part of the heterogeneous 3D Vision System-on-Chip which is analyzed in the case study. The system includes one mixed signal layer for the digitalization of the sampled images and one memory layer to temporary store the images. The components are connected via a 3D NoC. }\label{fig:case_study} 
\end{figure}

Thus, the  structure sketched in Fig.~\ref{fig:case_study} is analyzed.
In total seven different data/flit types are transmitted over the links: 1 for head-flits \textsf{h}, and 6 for body-flits (one per source) \textsf{s1}--\textsf{s6}.  
Each of the six image sensors in the MS layer is connected to one router R1--R6, which are connected by 2D links. Router R5 is connected via a 3D link with router R7 in the underlying MEM layer. Connected to R7 is a memory block to store the sensor data. Thus, over the links connecting R1$\rightarrow$R2, R3$\rightarrow$R2, R4$\rightarrow$R5 and R6$\rightarrow$R5 only data stemming from one sensor and head-flits are transmitted, resulting in unused virtual channels. Usage of virtual channels occurs in the links R2$\rightarrow$R5 and R5$\rightarrow$R7 as they are required for the transmission of data from 3 and 6 different sensors, respectively. In this section the same physical link structures as in Sec.~\ref{sec:eff} \& \ref{sec:sim_res} are considered. To obtain data flow matrices we run a simulation for the virtual channel-based and the virtual channel-less 3D NoC with our extended  simulator for a mean traffic injection rate of \unit[20]{\%} per sensor. To allow for subsequent bit-level simulations (to obtain reference values), the simulator is temporarily modified in a way that it saves the whole protocol of the transmitted flits. %which has a negative impact on the simulators memory usage.

%\subsection{Energy consumption}

After the simulations, the energy quantities per transmitted packet are determined with Python, employing: first our proposed high-level model, second the standard high-level model (neglects the effect of virtual channels), and third bit-level simulations. 
Thereby, we consider two separate traffic scenarios. In scenario 1 all six sensors capture road images with a resolution of 512$\times$512 pixels during the daylight, and in scenario 2 during the night.
We choose these particular traffic scenarios as they result in relatively high errors for our approach, which assumes that the cross-correlation between the individual data streams is zero. This is not
guaranteed if all sensors capture pictures of the same environment from different perspectives. 
The energy results with virtual channels are presented in the first row of Table~\ref{tab:case_strudy}. The NoC performance results, obtained by our simulator, are presented in the last row of the table. Energy consumption and network performance without virtual channels are shown in Table~\ref{tab:case_strudy1virtual channel}.
\begin{table}[]
	%\footnotesize
	\centering
	\caption{Link energy quantities and network performance with 4 virtual channels.}\label{tab:case_strudy}
	{\scriptsize\begin{tabular}{cccc}
		\toprule
		\multirow{2}{*}{\textbf{Data}} & \multicolumn{3}{c}{\textbf{Energy per transmitted packet [\unit{pJ}]}} \\ 
		&   Bit-level sim.         &  Presented model          & Standard model \cite{jafarzadeh2014data}                  \\ \cmidrule(lr){1-1}
		\cmidrule(lr){2-2} \cmidrule(lr){3-3} \cmidrule(lr){4-4}
		Uncoded &  4.18         &  4.15         & 2.40                  \\ %\cline{1-4}
		Gray &  4.11  (\textbf{\unit[-1.67]{\%}})        &  4.09 (\textbf{\unit[-1.44]{\%}})       & 2.23  (\textbf{\unit[-7.08]{\%}})                \\ %\cline{1-4}
		Corr &  2.69  (\textbf{\unit[-35.64]{\%}})        &  2.68    (\textbf{\unit[-35.42]{\%}})    & 2.71 (\textbf{\unit[+12.92]{\%}})                  \\ \midrule
		\multicolumn{2}{l}{Avg.\ flit latency: \unit[19.2]{ns}} & \multicolumn{2}{l}{Avg.\ network latency: \unit[105.4]{ns}}  \\ \bottomrule
	\end{tabular}}
\end{table}

\begin{table}[]
	%\footnotesize
	\centering
	\caption{Link energy quantities and network performance without virtual channels.}\label{tab:case_strudy1virtual channel}
	{\scriptsize\begin{tabular}{cccc}
			\toprule
			\multirow{2}{*}{\textbf{Data}} & \multicolumn{3}{c}{\textbf{Energy per transmitted packet [\unit{pJ}]}} \\ 
			&   Bit-level sim.         &  Presented model          & Standard model \cite{jafarzadeh2014data}                  \\ \cmidrule(lr){1-1}
			\cmidrule(lr){2-2} \cmidrule(lr){3-3} \cmidrule(lr){4-4}
			Uncoded &  2.39         &  2.40         & 2.40                  \\ %\cline{1-4}
			Gray &  2.22  (\textbf{\unit[-7.11]{\%}})        & 2.23 (\textbf{\unit[-7.08]{\%}})       & 2.23  (\textbf{\unit[-7.08]{\%}})                \\ %\cline{1-4}
			Corr &  2.70 (\textbf{\unit[+12.94]{\%}})        &  2.71    (\textbf{\unit[+12.92]{\%}})    & 2.71 (\textbf{\unit[+12.92]{\%}})                  \\ \midrule
			\multicolumn{2}{l}{Avg.\ flit latency: \unit[40.2]{ns}} & \multicolumn{2}{l}{Avg.\ network latency: \unit[193.3]{ns}}  \\ \bottomrule
	\end{tabular}}
\end{table}  
The simulation framework reveals that implementing virtual channels in the NoC allows to almost double the network performance for the analyzed application, as the average flit and network latency decreases by \unit[52.2]{\%} and \unit[45.5]{\%}, respectively. However, this performance gain is at an expense of a dramatic increase in link's power consumption of \unit[74.9]{\%} (beside increased complexity of the NoC architecture) .
Generally, in all analyzed realistic NoC traffic scenarios, our proposed model precisely predicts the energy consumption (error below \unit[1]{\%}, compared to results obtained by precise, but computational extensive, bit-level simulation).
In contrast, the previous high-level model which neglects the effect of virtual channels, leads to an error of almost \unit[50]{\%}, for the virtual channel-based NoC,
although in the analyzed virtual channel scenario less than \unit[50]{\%} of the links actually make use of their virtual channels.
However, the two links which use virtual channels show the highest energy consumption, and for these links the traditional model leads to an underestimation of the energy consumption by more than a factor of four.

After the simulation, we analyze the integration of two overhead free low-power coding approaches for the body-flits: correlator- and Gray-coding~\cite{5876688}. We analyze overhead free, low-complex, low-power codes as an induced bit-overhead would increase the buffer/memory cost. Additionally, both encoding techniques can be hidden in the AD converters of the sensors, to mitigate the implementation overhead.
Gray-encoding reduces the switching activities for a sequential transmission of the highly correlated pixels, while a correlator mainly affects the bit probabilities. As we know from the previous section, the correlator-coding shows good coding efficiency for multiplexed data streams, while the efficiency for no multiplexing is rather poor. The energy quantities for the transmission of the encoded data, instead of the RAW data, are also shown in Table~\ref{tab:case_strudy} (with virtual channels) and in Table~\ref{tab:case_strudy1virtual channel} (without virtual channels). The (estimated) energy reductions due to the coding approaches are shown in bold. Our presented model, in accordance with the bit-level simulations, indicates that the correlator leads to a far better coding efficiency (\unit[-36]{\%} instead of \unit[-1]{\%}). Thus, if virtual channels are integrated together with an end-to-end correlator coding, the link's energy consumption compared to the virtual channel-less system will be reduced from \unit[52.2]{\%} to \unit[12.5]{\%}.

In comparison, the standard model that neglects virtual channels, erroneously predicts a much higher coding efficiency for the Gray-coding and even a significant increase in the energy consumption (negative coding gain) for the correlator-coding. This underlines that, using any model other than ours for a high-level performance estimation, results in implementing inefficient coding techniques and a dramatic underestimation of the energy quantities.  

%\subsection{Buffer depth and number of virtual channels}

%\input{chapters/bufferDepth}

\section{Conclusion}
This work presents a NoC simulator with the first model to precisely estimate the data dependent dynamic energy consumption of 2D and 3D links even at the presence of virtual channels. The simulator is implemented in C++ and SystemC. Thus, it allows for an early design stage estimation of the NoC energy requirements. Furthermore, it enables the derivation and a fast evaluation of low-power data encoding techniques for links with virtual channels. In combination with proposed NoC simulator, a full design space exploration including link energy is possible for NoCs for the first time. The model shows negligible errors for realistic NoC traffic scenarios, with and without implemented coding techniques.

\section*{Acknowledgment}
This work is funded by the German Research Foundation (DFG) project GA 763/7-1 and PI 477/8-1.

\bibliography{biblio_patmos2018,biblio_moritz}

\end{document}